\documentclass[%
 reprint,
 preprintnumbers,
 amsmath,amssymb,
 aps,
]{revtex4-1}

\usepackage{graphicx}% Include figure files
\usepackage{dcolumn}% Align table columns on decimal point
\usepackage{bm}% bold math
\usepackage{subfigure}
\usepackage{color}
\usepackage{float}% Include figure files
\usepackage[normalem]{ulem}%underlining and strike-out in text mode
\usepackage{cancel} %cross-out in maths mode
\def\Pm{\mbox{\rm P}_M}
\def\Rm{\mbox{\rm R}_M}

\newcommand{\EQ}{\begin{equation}}
\newcommand{\EN}{\end{equation}}
\newcommand{\EQA}{\begin{eqnarray}}
\newcommand{\ENA}{\end{eqnarray}}

\newcommand{\Eq}[1]{Eq.~(\ref{#1})}

\newcommand{\Fig}[1]{Fig.~\ref{#1}}
\newcommand{\FFig}[1]{Figure~\ref{#1}}
\newcommand{\Figs}[2]{Figs.~\ref{#1} and \ref{#2}}

\newcommand{\mean}[1]{\overline #1}
{}
{}
{}
{}
\newcommand{\meanBB}{\overline{\mbox{\boldmath $B$}}{}}{}
\newcommand{\meanUU}{\overline{\mbox{\boldmath $U$}}{}}{}

\def\kf{k_{\rm f}}
\newcommand{\nab}{\mbox{\boldmath $\nabla$} {}}
\newcommand{\bb}{\bm{b}}
\newcommand{\ii}{{\rm i}}
\newcommand{\kk}{\bm{k}}
\newcommand{\xx}{\bm{x}}
\def\cs{c_{\rm s}}
\newcommand{\uu}{\mbox{\boldmath $u$} {}}
\newcommand{\BB}{\bm{B}}
\def\half{{\textstyle{1\over2}}}
\def\urms{u_{\rm rms}}

\newcommand{\meanB}{\overline{B}}
\newcommand{\meanu}{\overline{u}}

\def\apj{ApJ}

\def\mnras{MNRAS}
\def\aap{A\&A}
\def\apjl{ApJ}

\def\physrep{PhR}
\def\pre{PRE}

\def\nat{Natur}

\newcommand\zhetp{JETP}
\newcommand\jetp{Sov.\ Phys.\ JETP}
\newcommand{\EMF}{\mbox{\boldmath ${\cal E}$} {}}

\begin{document}

\preprint{NORDITA-2019-044}

\title{Efficient quasi-kinematic large-scale dynamo 
as the small-scale dynamo saturates}% 

\author{Pallavi Bhat}
\thanks{P.Bhat@leeds.ac.uk}
\affiliation{% 
Plasma Science and Fusion Center,
Massachusetts Institute of Technology, Cambridge, MA 02139, USA
}%
\affiliation{% 
Department of Applied Mathematics, 
University of Leeds, Leeds, LS2 9JT, UK 
}%

\author{Kandaswamy Subramanian}
\affiliation{IUCAA, Post Bag 4, Ganeshkhind, Pune 411007, India}

\author{Axel Brandenburg}
\affiliation{Nordita, KTH Royal Institute of Technology and Stockholm University,
SE-10691 Stockholm, Sweden}
\affiliation{Department of Astronomy, AlbaNova Center, Stockholm University, SE-10691 Stockholm, Sweden}
\affiliation{JILA and Laboratory for Atmospheric and Space Physics, University of Colorado, CO 80303, USA}
\affiliation{McWilliams Center for Cosmology \& Department of Physics, Carnegie Mellon University, PA 15213, USA}

\date{\today, $ $Revision: 1.63 $ $}

\begin{abstract}
Large-scale magnetic fields in stars and galaxies are thought to arise by mean-field dynamo action due to
the combined influence of both helical turbulence and shear. 
Those systems are also highly conducting and the turbulence therein leads to 
a fluctuation (or small-scale) dynamo which more rapidly amplifies magnetic field fluctuations on the eddy scales and smaller. Will this then
interfere with and suppress the mean (or large-scale) field growth?
Using direct numerical simulations of helical turbulence
(with and without shear),
we identify a novel quasi-kinematic large-scale dynamo which operates 
as the small-scale dynamo saturates.
Thus both dynamos operate efficiently, one after the other, and lead to the generation of significant large-scale fields. 
\end{abstract}

\maketitle

%\tableofcontents
Magnetic fields coherent on large-scales, larger than
the scales of turbulent motions in the system, are prevalent in stars and 
disk galaxies. Their origin is thought to lie in
mean-field or large-scale dynamo (LSD) 
action due to helical turbulence often combined with shear. 
Turbulence in stars and galaxies also has a very high magnetic Reynolds number $\Rm$.
This generically leads to a fluctuation or small-scale dynamo (SSD) which
grows random and small-scale magnetic fields more rapidly \cite{Kaz68,KA92}.
Here, small scales correspond to scales smaller than the outer scale of the
turbulence. A question of outstanding importance is whether these two dynamos
(LSD and SSD) cooperate or compete with each other
in the presence of rapidly growing fluctuations \citep{VC92}. Moreover, can we see
evidence of both dynamos in a large $\Rm$ system?

Earlier work using direct numerical simulations showed that 
helically forced turbulence at large $\Rm$ acts as a unified dynamo 
in the kinematic stage and grows magnetic
fields, of both large and small scales, with a shape-invariant 
eigenfunction \citep{SB14,BSB16}.
It is also known that saturation of the dynamo occurs first at small scales, and then at progressively larger scales. 
Thus, the field at scales larger than the forcing scale of the turbulence
continues to grow even when the field at smaller scales has already saturated \citep{BSB16}. 
Eventually, the large-scale field goes into the growth phase governed by the resistive decay of small-scale helicity \citep{B01}. 
In this work, we use helically forced turbulence with uniform shear
(and sometimes without)
to identify a novel intermediate second stage of exponentially growing large-scale field, 
different from that of the kinematic unified dynamo, 
due to a quasi-kinematic LSD. 
This arises in a previously unexplored parameter regime, but one that is in fact expected to be generic. 

We have run direct numerical simulations (DNS) 
of forced weakly compressible turbulence solving the
magnetohydrodynamics (MHD) equations in a periodic or
shearing-periodic cube of size $L^3$ with $L=2\pi$, so
the smallest wavenumber is $k_1=2\pi/L=1$.
We follow a setup similar to that of Ref.~\cite{BSB16},
except now we also have cases with uniform linear shear.
The simulations were performed with the \textsc{Pencil Code}
\footnote{DOI:10.5281/zenodo.2315093, github.com/pencil-code}, and
have a resolution of $512^3$ and a magnetic Prandtl number of $\Pm=10$.
We have specified the relevant parameters for each run in Table~\ref{runs}.
Turbulence is driven at the forcing wave number of $\kf/k_1=4$ (or $8$ in one case).
In our DNS, the unit of velocity is the sound speed $\cs$, and that of time is $(\cs k_1)^{-1}$. 
Our two main runs -- one with shear (Run~A) and one without shear (Run~B) --
have helical forcing which can lead to LSD action.
In the following, we discuss results from these runs and compare
with two similar runs (C and D) with
non-helical forcing and thus only SSD action. 

\begin{figure}[t!]
\includegraphics[width=0.48\textwidth]{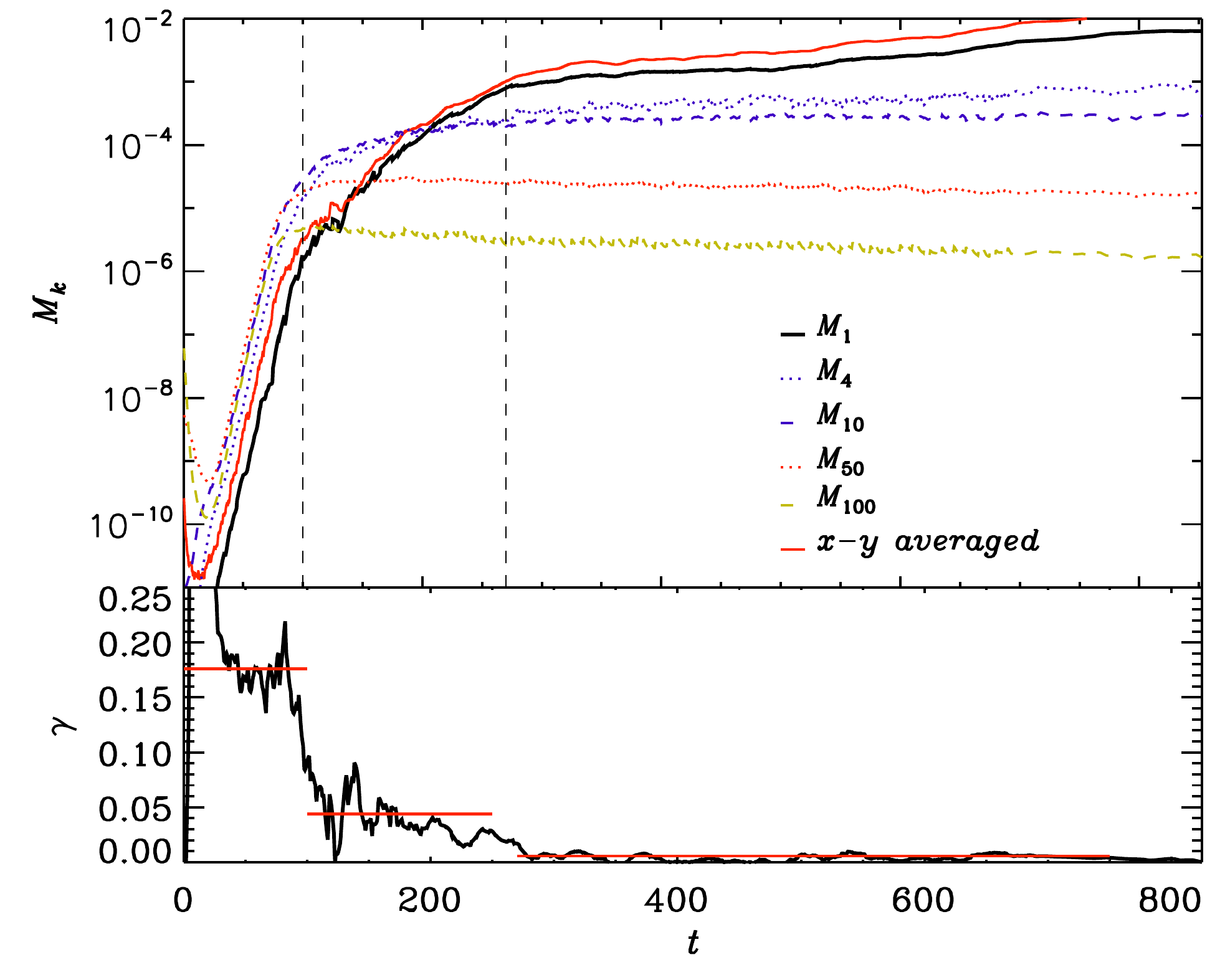}
\caption{Evolution of $M_k(t)$ for $k=1$, $4$, $10$, $50$, and $100$ for Run~A.
Evolution of energy in $xy$ averaged field $\meanBB_{\rm rms}^2$ in solid red.
The bottom panel shows $\gamma$, the growth rate of $M_1$.}
\label{spectimeRunA}
\end{figure}

In the top panels of \Figs{spectimeRunA}{spectimeRunB}, 
we show the evolution of the magnetic energy spectrum $M_k$ at certain wave numbers $k$.
We characterize large- and small-scale fields through the magnetic energy at $k<\kf$
and $k>\kf$, respectively. The quantity
$M_1$ is seen to grow exponentially at the same kinematic rate as others until about $t=100$. 
However, there is a second phase between $t=100$ to $t\sim250$--$270$, 
where $M_1$ grows exponentially at a different, albeit slower rate. 
Meanwhile all modes with $k\ge\kf$ have slowed down towards saturation.
In a third phase, after $t\sim 300$, the growth of $M_1$ 
further slows down to a resistively limited rate.
For Run A, we have also calculated the large-scale field $\meanBB_{\rm rms}$ from 
horizontal or $xy$ averaging. 
In the top panel of \Fig{spectimeRunA}, we show that evolution of $\meanBB_{\rm rms}^2$ closely follows the $M_1$ curve. 
Thus, the three different phases 
of growth are observed also in the evolution of the large-scale field calculated from this different method. 
In the bottom panels of \Figs{spectimeRunA}{spectimeRunB},
we plot the growth rate of $M_1$,
defined as $\gamma=d \ln(M_1)/dt$. 
Three stages of growth can clearly be identified.

We propose that this second distinctive phase of growth of the large-scale field is in fact
a standard LSD -- as predicted by mean-field dynamo theory.
In the first stage, which is entirely kinematic, the faster SSD 
is the main driver and thus governs the growth rate of magnetic energy over all scales. 
The large-scale field at $k=1$ grows as a low wave number tail of the SSD eigenfunction,
coupled to the small scales, possibly through a $k^{3/2}$ \cite{Kaz68,KA92}
or a $k^{7/6}$ \cite{SB14} slope.
Once the growth at smaller scales has slowed down,
large-scale dynamo action becomes prominent.

\begin{figure}[t!]
\includegraphics[width=0.48\textwidth]{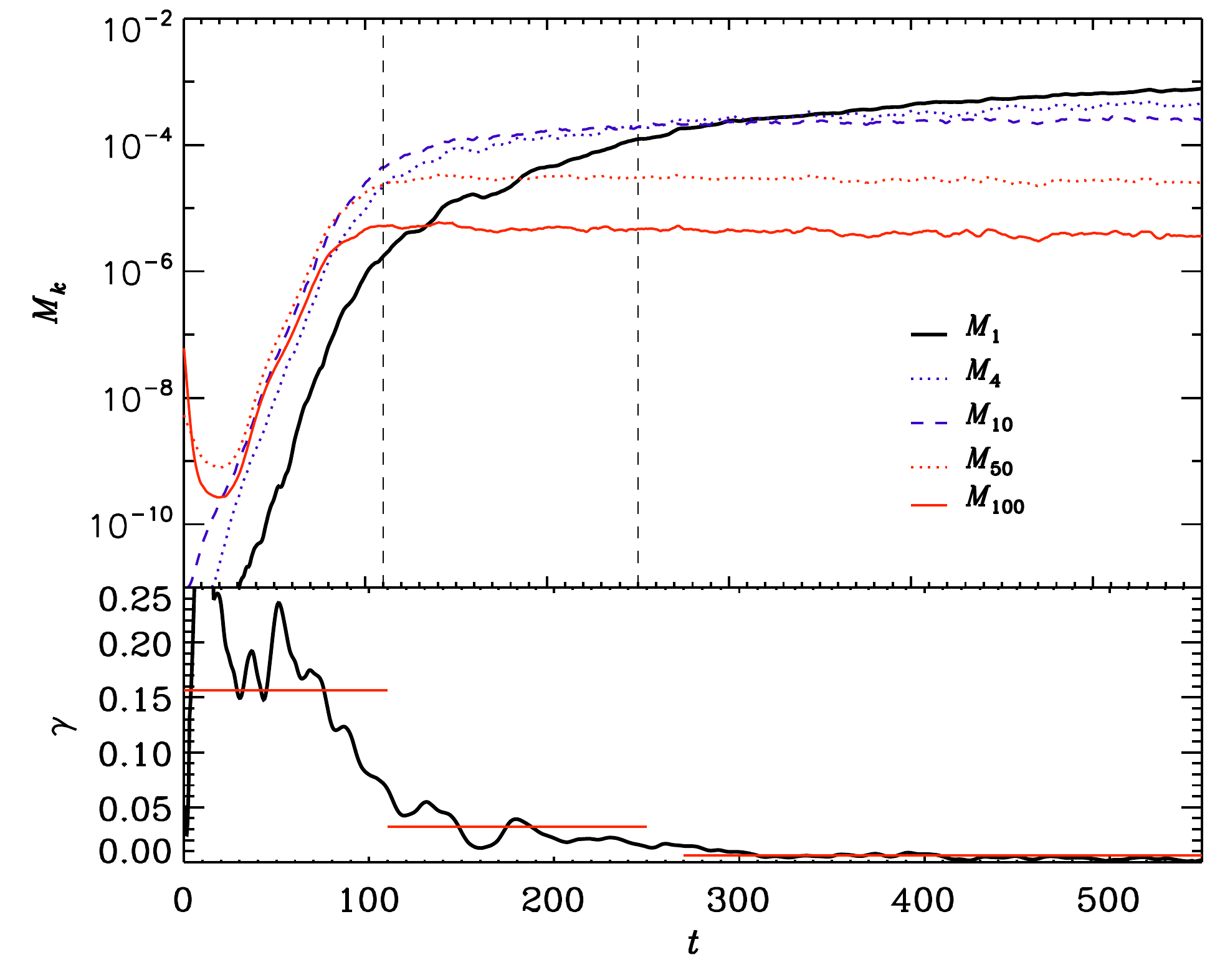}
\caption{Evolution of $M_k(t)$ for $k=1$, $4$, $10$, $50$, and $100$ has been shown for Run~B.}
\label{spectimeRunB}
\end{figure}

For the following discussion, it is helpful to
refer to the standard mean-field equations
obtained by splitting the induction equation into one
for the mean or large-scale field $\meanBB$
and the fluctuating or small-scale field $\bb$ \citep{Mof78},
\begin{equation}
\label{meaninduction}
\frac{\partial \meanBB}{\partial t} = \nab \times \left( \meanUU \times \meanBB
+ \EMF- \eta \bf \nab \times \meanBB \right),
\end{equation}
\begin{equation}
\label{residual}
\frac{\partial \bb}{\partial t}={\bf \nab}\times\left(\meanUU\times \bb+
\uu\times \meanBB-\eta \bf \nab \times \bb+\bm{G}\right).
\end{equation}
Here $\EMF = \overline{\uu \times\bb}$ and
$\bm{G} = \uu\times \bb-\EMF$ is a term nonlinear in the fluctuations. 
The mean velocity accounts for the linear shear, $\meanUU=(0,Sx,0)$, 
where $S=\mbox{const}$. 
We also find that helical forcing combined with 
shear induces 
a large-scale flow $\meanUU = (\mean{U}_x(z), \mean{U}_y(z), 0)$ on $xy$ averaging but this 
does not appear to affect the LSD (see below).
This phenomenon is the vorticity dynamo \citep{EKR03,KMB09},
which is known to be suppressed by the magnetic field \citep{KB09,GHJ17}.

\begin{figure}[t!]
\includegraphics[width=0.48\textwidth]{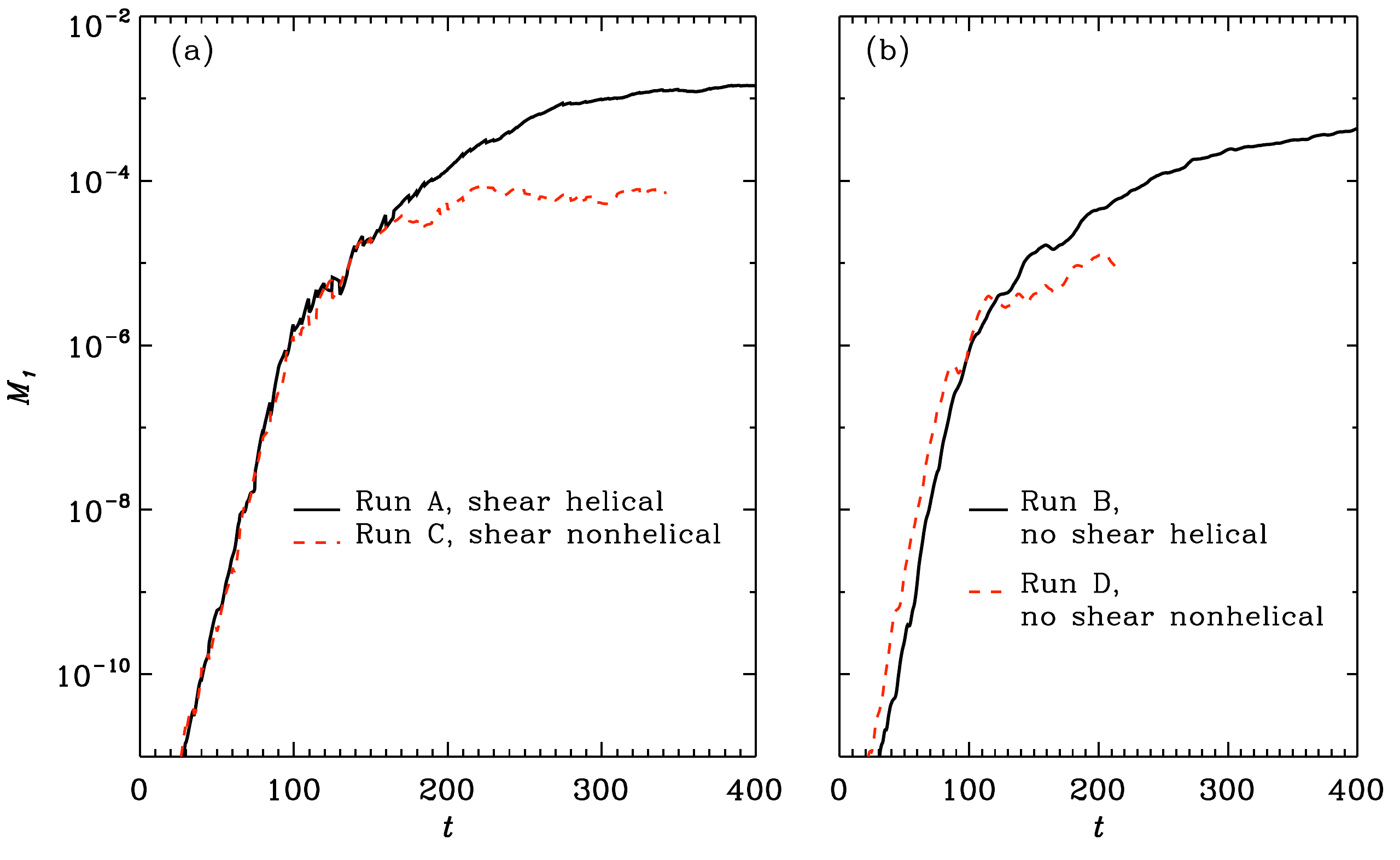}
\caption{Comparison of $M_1$ curves between Runs~A and C in the panel (a) and between Runs~B and D in the panel (b).} 
\label{m1comp}
\end{figure}

During the kinematic stage, the small-scale field is mainly driven by
$\nab \times \left(\uu \times \bb \right)$, which leads to SSD action. 
All the terms that depend on averaged or mean quantities 
are not significant initially.
We have checked that the shear in our simulations is small enough that
its effect on SSD growth is unimportant \cite{Nishant}.
The small-scale field then grows exponentially as 
${\bb}={\bb}_0 \exp{(\gamma_{\rm SSD} t)}$, 
where $\gamma_{\rm SSD}$ is the SSD growth rate.
In \Eq{meaninduction}, the time evolution of $\EMF$, which drives $\meanBB$,
is then controlled by 
the exponentially growing low wave number tail of the small-scale field, given that 
the velocity field ${\uu}$ is in statistical steady state. 
The shear term $\nab \times \left( \meanUU \times \meanBB \right)$
is subdominant as $\meanBB$ is at this stage much smaller than $\bb$.  
The rate of change of $\meanBB$ is therefore
expected to be nearly the same as $\gamma_{\rm SSD}$.
This can be seen also in \Fig{m1comp}, where in the left panel we show that the time 
evolution of $M_1$ in both the helical shear dynamo simulations
(Run A) and non-helical shear dynamo (Run C) coincide in the kinematic stage. 
Similarly, in the right panel of \Fig{m1comp}, the $M_1$ curves 
from  helical dynamo (Run B) and non-helical dynamo (Run D) coincide.
Thus, the kinematic stage is primarily driven by the SSD with $\meanBB$ being
enslaved to $\bb$. 
 
\begin{figure}[t!]
\includegraphics[width=0.48\textwidth]{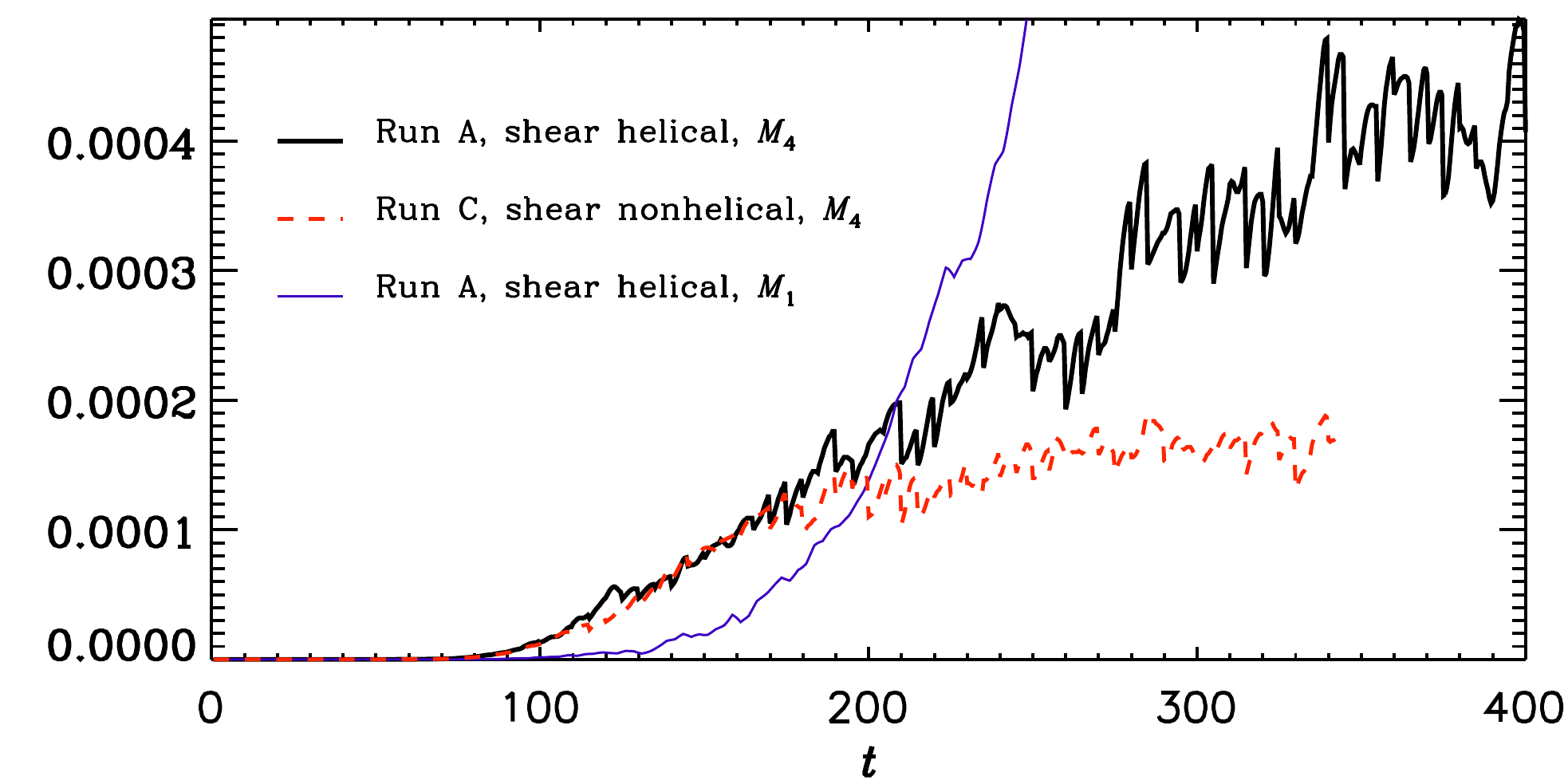}
\caption{Comparison of $M_4(t)$ shown for Run~A and Run~C.
Also comparison of curves from Run~A -- $M_1$ growing exponentially while $M_4$ grows linearly.}
\label{sscompare}
\end{figure}

Eventually, the growth of the small-scale field slows down from an exponential 
to a more linear 
form as the SSD begins to saturate. 
We find this coincides with a second stage of exponential growth of the large-scale field.
As seen in the upper panel of \Fig{spectimeRunA}, the
evolution of $M_4$ and field at even smaller scales $M_{10}$,
slow down at around $t=100$, when $M_1$ switches to a different rate of exponential growth. 
\FFig{sscompare} shows explicitly the exponentially growing $M_1$ versus linearly growing $M_4$, from Run~A.
We can understand the second stage of growth of the large-scale field in the following manner.
In \Eq{residual}, besides the shear and SSD terms, now there are contributions due to the terms containing mean quantities. 
In particular, the term $\nab \times (\uu\times \meanBB)$, interpreted as the tangling of the large-scale 
field, is expected to be responsible for additional growth of small-scale fields
over and above that when there is no large-scale dynamo. 
We show that this is indeed the case in \Fig{sscompare}, where 
the linear growth rate of the $M_4$ in the helical dynamo in Run~A
is larger than that of non-helical dynamo (Run~C) by a factor of about 8.

\begin{table}
\caption{
Summary of all the runs.}
\vspace{10pt}
{\begin{tabular}{lccccccc}
\hline
\hline
Run & $\kf$ & $u_{\rm rms}$  & $S\tau$ & forcing & $\Rm$ & $\gamma_{\rm theo}$ & $\gamma_{\rm meas}$\\
\hline
A & 4 & 0.13 & 0.38 & helical & 812 & 0.055 & 0.036\\
B & 4 & 0.18 & 0    & helical & 1062 & 0.060 & 0.032\\
C & 4 & 0.13 & 0.38 & non-hel & 812 & -- & -- \\
D & 4 & 0.19 & 0    & non-hel & 1187 & -- & -- \\
E & 4 & 0.10 & 0.25 & helical & 1000 & 0.033 & 0.026\\
F & 4 & 0.09 & 0.11 & helical & 812  & 0.017 & 0.013\\
G & 8 & 0.09 & 0.27 & helical & 281 & 0.051 & 0.032\\
H & 4 & 0.18 & 0  & helical & 531 & 0.06 & 0.034\\
\hline
\label{runs}\end{tabular}}
\end{table}

At this stage, as the tangling of large-scale field by the random velocity $\uu$
becomes the more dominant mechanism for growth of small-scale fields $\bb$,
this leads to a correlation between $\bb$ and $\uu$, 
proportional to $\meanBB$. The emf $\EMF = \overline{\uu\times\bb}$,
which then depends on $\meanBB$, can be estimated by  
the usual closure scheme, first order smoothing approximation (FOSA), 
to be $\EMF = \alpha \meanBB - \eta_{\rm t}\nab \times \meanBB$ \citep{Mof78}.
Here $\alpha$ and $\eta_{\rm t}$ are the 
turbulent transport coefficients determining the 
effect of small-scale turbulence on the large-scale magnetic field.
Thus, \Eq{meaninduction} for the large-scale field transforms to,
\EQ
{\partial\meanBB \over \partial t}=
\nab \times \left(\meanUU\times\meanBB + \alpha\meanBB\right)
+\eta_{\rm T}\nabla^2\meanBB,
\quad\nab\cdot\meanBB=0,
\label{mfd}
\EN
with $\eta_{\rm T}=\eta+\eta_{\rm t}$.
This is the standard mean-field dynamo equation,
which has solutions in a periodic box of the form
$\meanBB(\xx,t)=
\mbox{Re}\left[\hat{\BB}(\kk)\,\exp(\ii\kk\cdot\xx+\lambda t)\right]$
at the kinematic stage.
For simplicity assume the large-scale field varies only along $z$ and
so $k_y=k_x=0$. Then the eigenvalue 
$\lambda$ is given by \citep{BS05},
\EQ
\lambda_\pm=-\eta_{\rm T} k_z^2\pm(\alpha^2 k_z^2-\ii\alpha Sk_z)^{1/2}.
\EN
We note that the $\meanUU$ from the vorticity dynamo does not affect the dispersion
relation as $\nab\times(\meanUU\times\meanBB)=0$ when
$\mean{B}_z=\mean{U}_z=0$ and the fields 
depend only on $z$.
For Run~A, the vorticity dynamo becomes suppressed as the magnetic
field continues to saturate \footnote{See Supplemental Material,
which includes Refs.~[10-11], regarding the suppression of the
vorticity dynamo by the magnetic field.}.

For the case without shear ($\alpha^2$ dynamo), the growing mode
has $\lambda = |\alpha|k_z - \eta_{\rm T} k_z^2$. 
When shear dominates (standard $\alpha\Omega$ dynamo), such that
$\alpha k_z/S\ll1$,
\EQ
\mbox{Re}\lambda_\pm\approx-\eta_{\rm T} k_z^2\pm|\half\alpha Sk_z|^{1/2},
\label{lam_approx}
\EN
\EQ
\mbox{Im}\lambda_\pm\equiv-\omega_{\rm cyc}\approx\pm|\half\alpha Sk_z|^{1/2}.
\label{om_approx}
\EN

\begin{figure}[t!]
\includegraphics[width=0.4\textwidth]{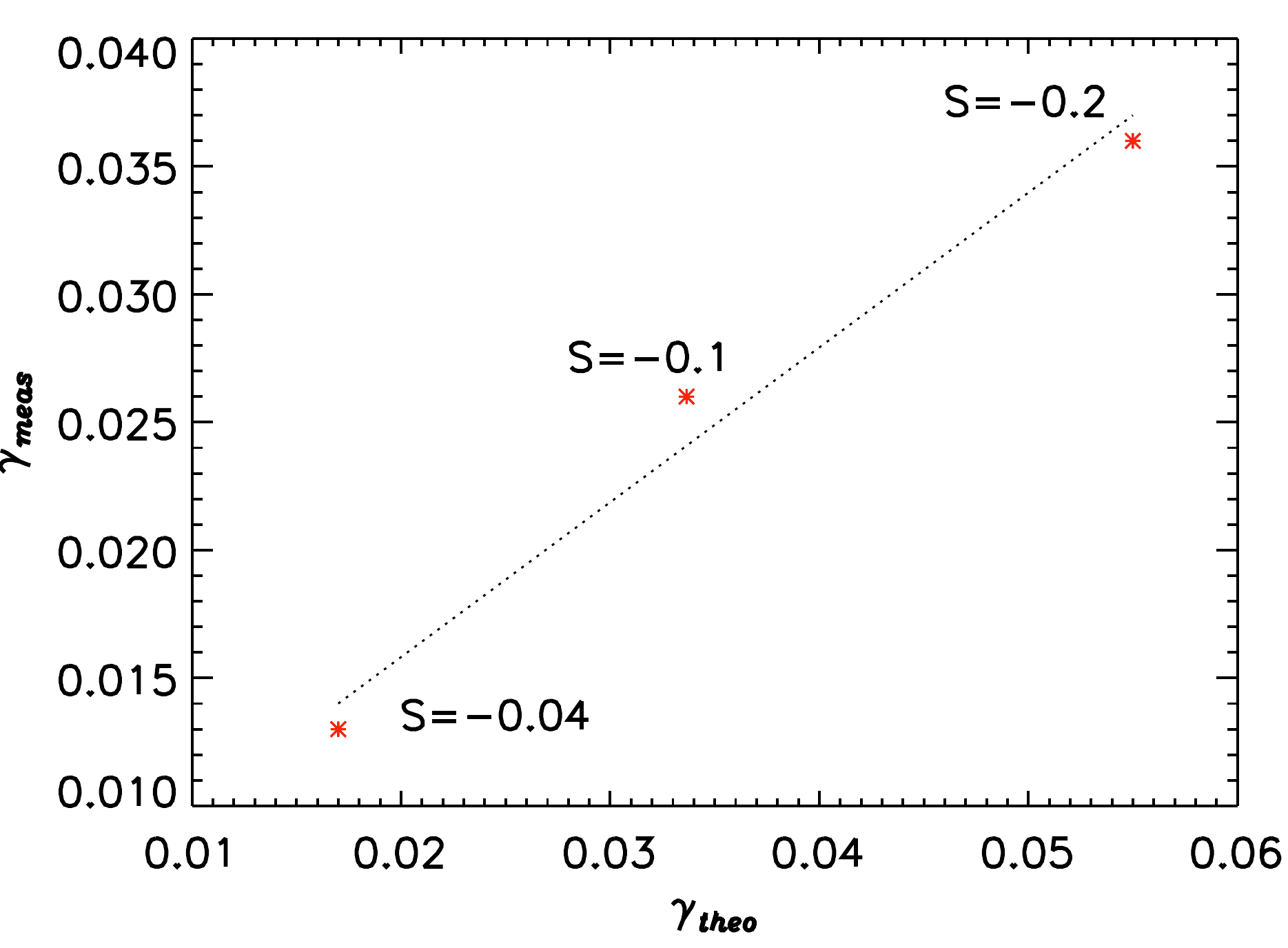}
\caption{Comparison between
theoretical $\gamma_{\rm theo}=-\eta_{\rm T} k^2\pm|\half\alpha Sk_z|^{1/2}$ and
measured $\gamma_{\rm meas}=d\ln{M_1}/dt$, for Runs~A, E and F (with varying shear, $S$).}
\label{varshear}
\end{figure}

We now ask, can the second stage
of exponential growth of $\meanBB$ shown in \Fig{spectimeRunA},
be understood in terms of the above standard mean-field dynamo properties?
For homogeneous, isotropic and fully helical turbulence forced
at a wave number $\kf$, we estimate
$\alpha \sim u_{\rm rms}/3$ and $\eta_t \sim u_{\rm rms}/(3\kf)$
\citep{SBS08}. The second term (with shear) in \Eq{lam_approx} governs 
the growth rate, while the first term would be smaller in the supercritical case. 
In our Run A, $\urms\sim0.13$, $\kf=4$ and $S\tau\sim0.38$, which leads to
$\gamma_{\rm theo}=\mbox{Re}\lambda_+\sim0.055$, 
which is larger than the measured value of $\gamma_{\rm meas}\sim0.036$.
This yields an `efficiency factor' $c_{\rm effic}\sim 0.65$.
Note that in the runs with shear, 
we estimate the $\urms$ after subtracting out $\meanUU$.

From \Eq{lam_approx}, we observe that 
the growth rate is not expected to change much as we change $\kf$. 
We have run a case with $\kf=8$, where $\urms\sim 0.09$ and $S\tau\sim 0.27$,
yielding $\gamma_{\rm theo}\sim 0.051$. This theoretical estimate is 
similar to the $\gamma_{\rm theo}$ of Run~A.
This is also confirmed by the measurement of the growth rate of $\sim0.032$
(similar to measured value of $0.036$ in Run~A). 
In the no-shear case, the theoretical growth rate estimate is given by $\urms\kf/12$.
For Run B, where $\urms\sim0.18$ this leads to $\gamma_{\rm theo}\sim0.06$. 
Here, with $\gamma_{\rm meas}\sim0.032$, we have $c_{\rm effic}\approx 0.53$.

We have varied the shear parameter $S\tau$ to see its effect
on this second stage growth rate of the large-scale field. 
In \Fig{varshear}, we compare the theoretical estimate of the growth rate 
for the Runs~A, E and F (with different values of the shear parameter) against the measured value.
We find that $c_{\rm effic}$ is roughly the same in all three cases, thus leading 
to the points in \Fig{varshear} falling nearly on a straight line. 

\begin{figure}[t!]
\includegraphics[width=0.4\textwidth]{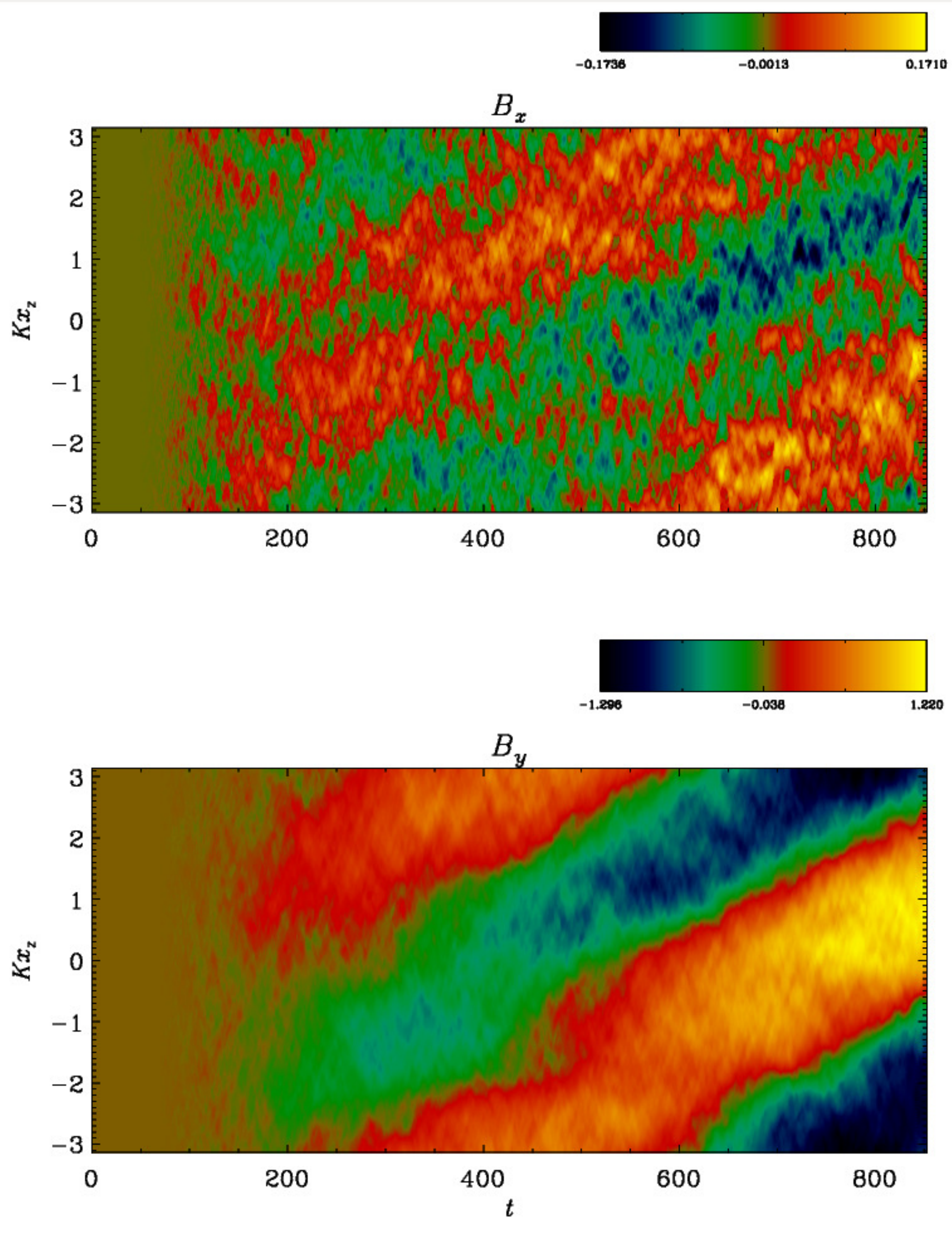}
\caption{Space-time diagram where the field components have been $xy$ averaged.}
\label{butter}
\end{figure}

Next we examine the oscillatory behavior of the LSD in the runs with shear. 
In \Fig{butter}, we show $xy$ averaged fields $\meanB_x$ and $\meanB_y$ in a $zt$ space-time diagram. 
As in earlier work at lower $\Rm$ \citep{KB09},
the oscillations begin only during the second stage of growth of the large-scale field.
We make a theoretical estimate of the time period of this cycle 
from the mean-field theory. If we take 
$\mbox{Re}\lambda_+$ in \Eq{lam_approx}
to $0$ (which 
approximately holds as the LSD saturates), 
we obtain $\omega_{\rm cyc}=\eta_{\rm T} k_z^2$. 
Thus, the time period can written as $T=2\pi/\omega_{\rm cyc}=6\pi(\kf/k_1)^2$.
For Run A, such an estimate yields $T\sim300$ and we find 
from the simulation shown in \Fig{butter}, this is roughly the time period of the oscillations in the large-scale field. 
From these results, it appears that the mean-field theory is satisfactorily applicable 
to understand the LSD in this second stage. 

An important question is whether there are any effects due to varying $\Rm$ on the large-scale field growth in this stage?
To check this we have lowered in Run~H the $\Rm$ by a factor of $2$ compared to that of Run~B. 
We find that the growth rate in this Run~H is indeed the same as that in Run~B.
Lastly, the third stage of slow growth of the large-scale field is governed by
the resistive decay of small-scale magnetic helicity,
which had built up to suppress the $\alpha$-effect,
as shown by previous works \cite{B01,BBS02,BB02,BS05};
see also \footnote{See Supplemental Material for
details of the evolution of the large and small-scale helicity}.
At late times, there could also be a fourth phase of nonlinear 
mode switching of the helical dynamo when linear shear is included \cite{HRB11}; 
see also \footnote{See Supplemental Material,
which includes Ref.~[20], regarding a demonstration of
mode switching.}.

Our earlier simulation of the helical LSD \cite{BSB16},
also at $\Pm=10$, had a higher resolution ($1024^3$)
but a smaller $\urms\sim0.12$. Thus, the resulting growth rate 
$\urms\kf/12$ of the LSD was small
and thus it was difficult to delineate the second stage from the third stage
where growth is governed by resistive effects. 
It is important to recognize that our simulations are in a parameter regime which allows 
for separation in large-scale field growth time scales between the three stages. This
made it possible to identify the second stage here.
Also such a separation in growth time scales can be expected in astrophysical systems. 
A suggestion of an intermediate stage of LSD growth was made earlier \cite{KKB08}, 
but did not receive much emphasis then. 
Also, this was a very different system with convection and
boundaries permitting a magnetic helicity flux.

A caveat in our estimates is that we are applying mean-field theory
to a system which is already affected by the Lorentz force (as the SSD slows down) 
and thus the theory needs to include the nonlinear effects.
However, it may be possible that the nonlinearity affects only the small-scale fields at this stage,
while the theory for the large-scale field, involving an effective 
$\alpha$ effect could still be applicable and thus we
term it a quasi-kinematic LSD.
It has been suggested that the spatial 
intermittency of the magnetic field when SSD saturates could
still allow the LSD to operate \cite{S98}.
Note that the quasi-kinematic LSD here arises upon saturation of the SSD
and seems to be in alignment with the `suppression principle' 
put forward by \citet{TC2013},
although there the SSD is suppressed even in the kinematic stage due to shear \cite{NPCT17}.

In conclusion, we have demonstrated via direct numerical simulations the presence of
a novel second stage growth of large-scale field, one that occurs 
between the kinematic stage driven by the SSD and the saturation stage driven by magnetic helicity decay.
The SSD provides a seed for the LSD as postulated previously 
\cite{BPSS94}.
Interestingly, we find that this second stage of large-scale field 
growth sets in when the SSD
slows down.
Our detailed analysis reveals the important result that
the classical mean-field theory applies well to understanding the large-scale field characteristics
including its growth rate and frequency of oscillations. 
Moreover, the growth of the large-scale field in this second stage is independent of $\Rm$.
A major difficulty in the application of dynamo theory to large $\Rm$ 
systems like galaxies has been whether large-scale fields can grow
at all in the presence of the SSD?
However, our result now shows that exponential growth of fields by SSD is followed by an exponential growth of the large-scale field by LSD, thus indeed making it feasible 
to obtain large-scale fields on fast dynamical time-scales. Our work thus gives the first detailed evidence for how a large-scale dynamo
operates quasi-kinematically once the small-scale field has saturated and develops an identity even amidst strong small-scale fields. 

\acknowledgements
We thank Anvar Shukurov and Steve Tobias for helpful suggestions and discussions. 
PB thanks Nuno F.\ Loureiro for support under award number DE-SC0016215 from NSF-DOE.
This project was completed using funding from the European Research Council (ERC) under the 
European Union's Horizon 2020 research and innovation programme (grant agreement no.\ D5S-DLV-786780).
This work was supported through the National Science Foundation, grant AAG-1615100,
The simulations in the paper were performed on the MIT-PSFC partition
of the Engaging cluster at the MGHPCC facility (www.mghpcc.org), which
is funded by DoE grant number DE-FG02-91-ER54109.
Additional simulations were performed using resources provided by
the Swedish National Infrastructure for Computing (SNIC)
at the Royal Institute of Technology in Stockholm.

\bibliography{reftwody}

\newpage
\LARGE
\noindent
Supplemental Material

\large
\vspace{1mm}
\noindent
to ``Efficient quasi-kinematic large-scale dynamo as the small-scale dynamo saturates''

\normalsize
\vspace{1mm}
\noindent
by P.\ Bhat, K.\ Subramanian \& A.\ Brandenburg
%\\ \scriptsize{\today,~ $ $Revision: 1.63 $ $}\normalsize

%\vspace{1mm}
%\noindent

\subsection{Shear helical dynamo: large-scale flows and late time behavior}

\begin{figure}[b]
\includegraphics[width=\columnwidth]{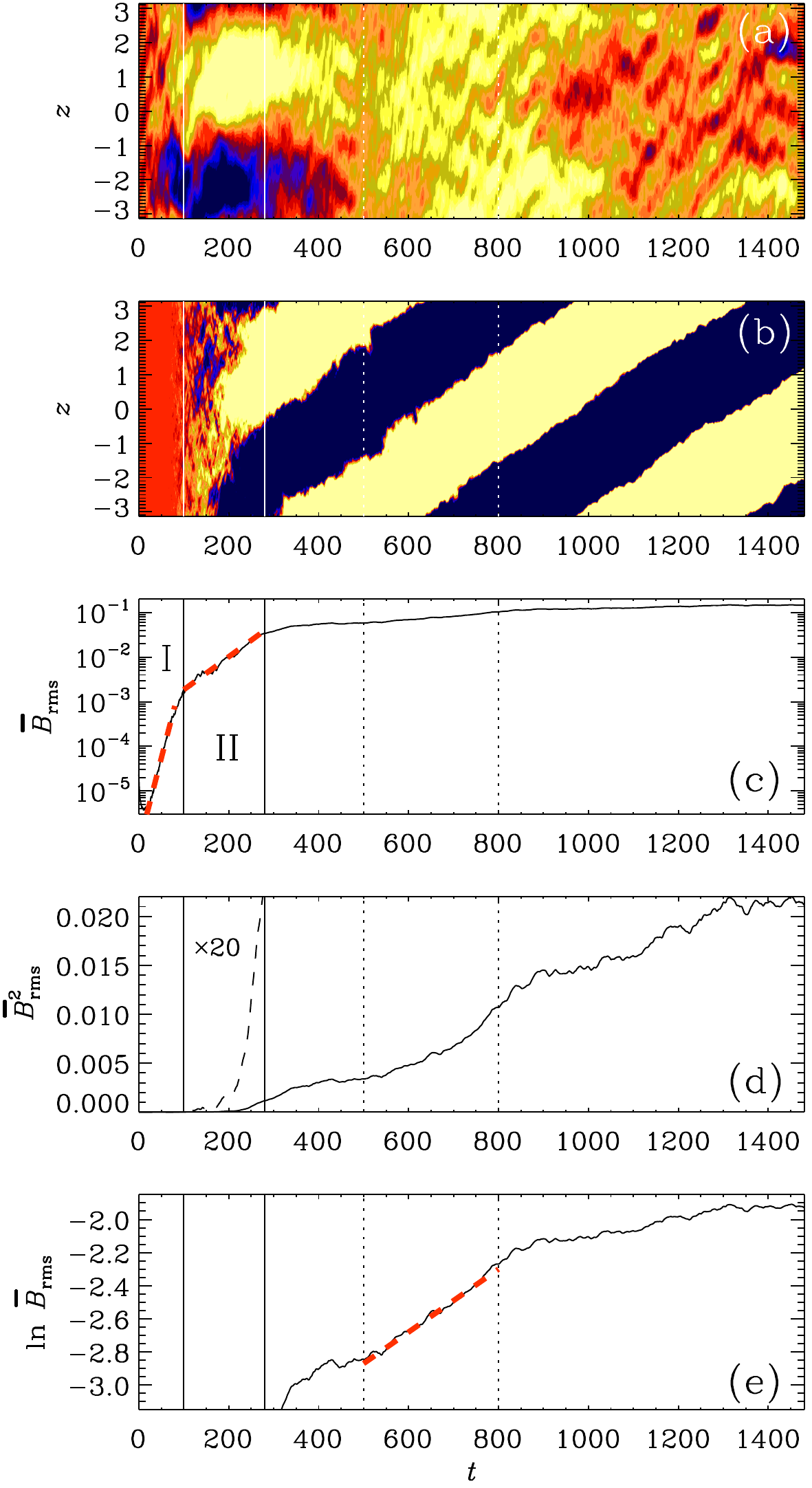}
\caption{Comparison of $\meanu_y$, $\meanB_y$, and three
representations of $\meanBB_{\rm rms}$ for Run~A.
The dotted line is scaled by $\times20$.}
\label{psat}
\end{figure}

\begin{figure}[t]
\includegraphics[width=\columnwidth]{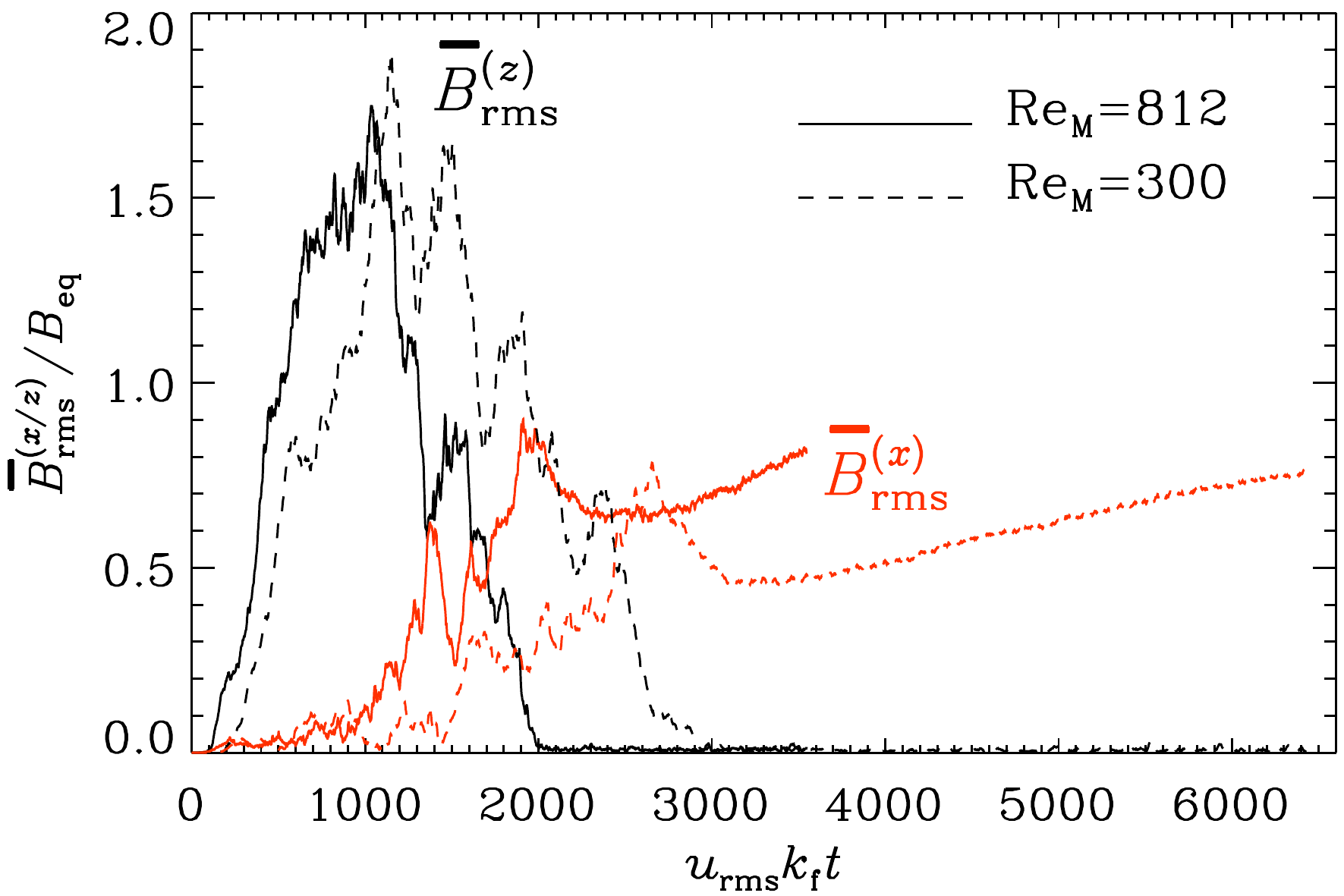}
\caption{Comparison of the rms values of the $z$ dependent ($xy$-averaged,
black lines) and $x$ dependent ($yz$-averaged, red lines) mean fields,
$\meanBB_{\rm rms}^{(z)}$ and $\meanBB_{\rm rms}^{(x)}$, respectively,
for Run~A (solid line) and a similar one at $\Rm=300$ (dotted lines).}
\label{pcomp2}
\end{figure}

We mentioned in the main paper that a large-scale component arises
in the velocity field.
This is shown in \Fig{psat}(a).
The large-scale flow is seen to arise early on, just after the end of
phase~I, at $t=100$.
This is due to a vorticity dynamo \cite{EKR03} and arises also without
magnetic field \cite{KMB09}.
Note that, because the large-scale flow varies only along the $z$ direction,
the relevant terms in the mean-field dynamo equation basically go to
zero and thus this large-scale flow is not responsible for any magnetic
field generation.
At $t=400$, the large-scale flow weakens, changes its form, and
disappears by $t=1000$.
This suppression is due to the magnetic field \cite{KB09,GHJ17}.
\FFig{psat}(b) shows that a dynamo wave begins to emerge at $t=100$,
but it has initially a larger phase speed than at later times.
In \Fig{psat}(c), we see that there is exponential growth in phases
I and II with different growth rates.
\FFig{psat}(d) shows that this growth of $\meanBB_{\rm rms}^2$ does not
grow linearly in time.
There is also evidence for another period of exponential growth 
at later times after phase II; see \Fig{psat}(e).
The growth rate however is very slow
and this perhaps arises due to a temporary fluctuation 
during the resistive phase.

\begin{figure*}[tp]
  \centering
\includegraphics[width=0.32\textwidth]{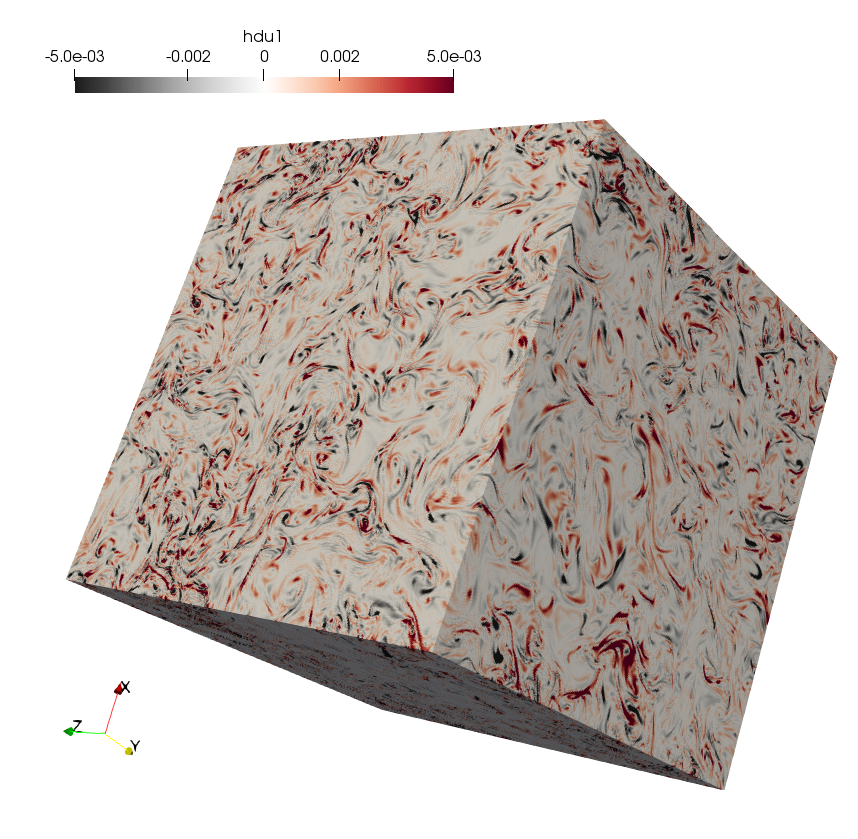}
\includegraphics[width=0.32\textwidth]{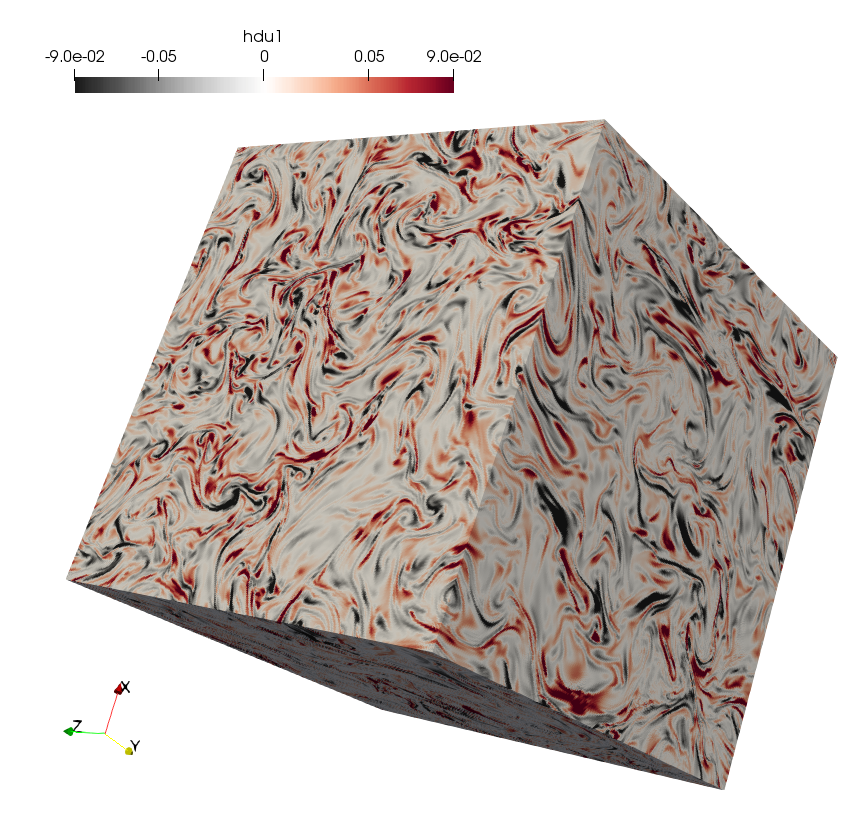}
\includegraphics[width=0.32\textwidth]{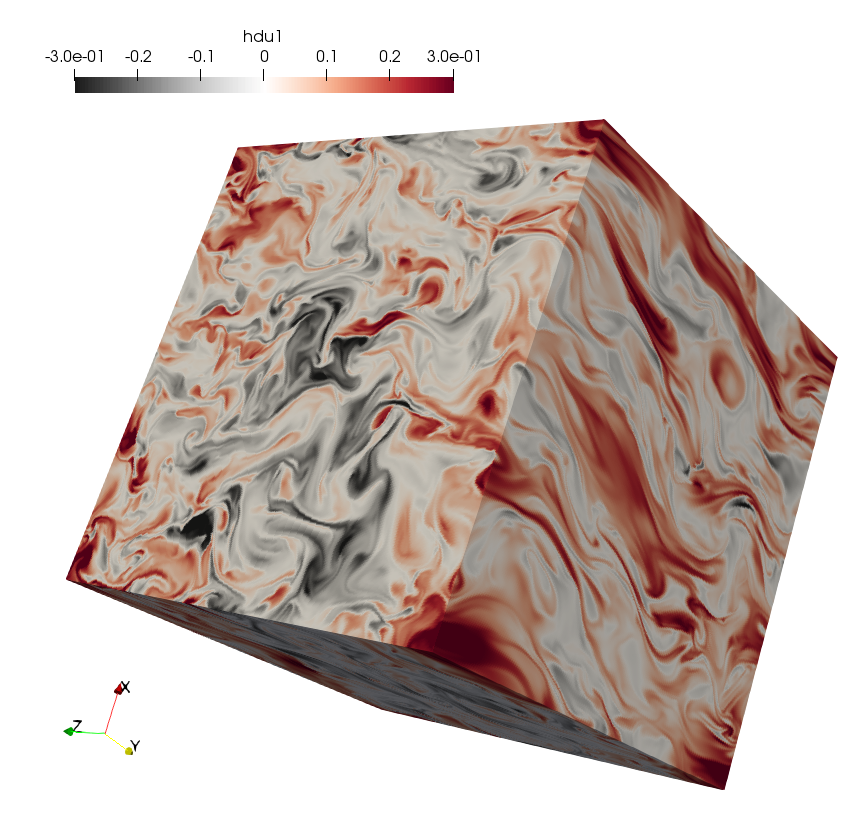}
  \caption{
$B_y$ for Run~A at $t=50$, $200$ and $400$.
   } 
  \label{byboxes}
\end{figure*}

As mentioned in the main paper, a fourth stage involves 
fratricide of this $\alpha\Omega$ dynamo by its $\alpha^2$ sibling \cite{HRB11}.
This can be seen by plotting the $x$ dependent ($yz$-averaged) mean field,
which we denote by $\meanBB_{\rm rms}^{(x)}$ with superscript ${(x)}$.
The usual $z$ dependent ($xy$-averaged) mean field is now denoted by
$\meanBB_{\rm rms}^{(z)}$ with superscript ${(z)}$.
This fratricide happens at much later times during $\urms\kf t=1000$--$2000$; see \Fig{pcomp2},
where the black solid curve of $\meanBB_{\rm rms}^{(z)}$ shows decay
and simultaneously, $\meanBB_{\rm rms}^{(x)}$ in solid red is growing. 
Importantly, it appears that the phase II is not $\Rm$ dependent
as seen by comparing the near parallel solid and dashed black lines in \Fig{pcomp2} between $t=100$ to $400$.
Thus, there is no evidence for catastrophic quenching of this phase in this range of $\Rm$ explored.
\subsection{Spatial organization of the field}

To understand how the magnetic field changes in different stages of
growth, we show in \Fig{byboxes} the magnetic field component $B_y$
of Run~A at times $t=50$, $200$ and $400$.
At $t=50$, the system is in kinematic stage and thus the fields are of small-scale nature. 
At $t=200$, the large-scale dynamo (LSD) is active and one finds that
the fields have started ordering themselves on larger and larger scales. 
In the third stage of resistive decay of helicity, the fields become
increasingly organized and at $t=400$, there is a coherent field
on the largest scale in the box.

\subsection{Magnetic helicity evolution in the dynamo with 
shear and helical forcing}

In the main paper we mentioned that 
the build up of small-scale helicity during the second stage is expected
to eventually quench the LSD.
Initially, the LSD due to $\alpha$-effect results in a helical 
polarization of the field, which is 
of opposite signs on small and large scales. The
Lorentz force associated with this small scale helical field
can back react to then quench the $\alpha$-effect 
\cite{B01,BBS02,BB02,BS05}.
Moreover, the large-scale field growth, which as discussed in the
main paper is similar in helical
and non-helical runs during the first stage when the SSD dominates,
can however be differentiated at later stages by its helicity properties. 
Thus, it is of interest to examine the magnetic helicity power spectrum
$H(k)$ for this standard signature of the LSD and 
study how it evolves in the first two stages of large-scale field growth discussed in the main paper.

\begin{figure}[h]
\includegraphics[width=0.48\textwidth]{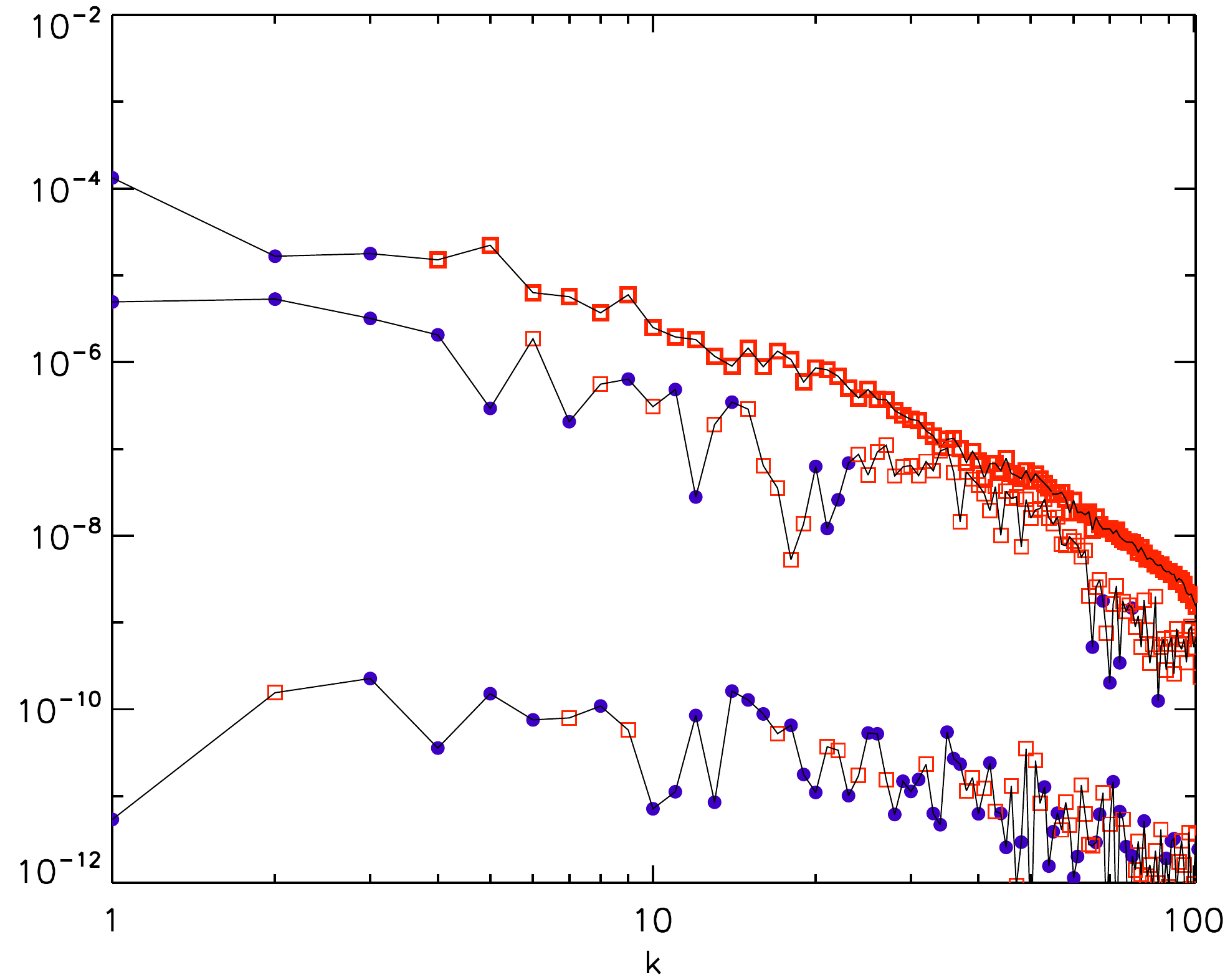}
\caption{The magnetic helicity power spectrum is shown at three times,
$t=100,150$ \& $250$ for Run~A. The blue diamonds represent negative 
helicity and red squares represent positive helicity.}
\label{helspec}
\end{figure}

In the kinematic stage, when the SSD is dominant, there is no clear
separation between the positively helical and the negatively helical
fields as shown by $H(k)$ in \Fig{helspec} for Run~A.
However, towards the end of the second stage (the curve at $t=250$), a
clear separation in scales based on helicity develops, i.e., the helicity
on smaller $k$, $k<\kf$ is one sign represented by blue diamonds and the
helicity on larger $k$ is the opposite sign represented by red squares.
It is this accumulation of small-scale helicity that could induce
a magnetic back reaction to the initial kinetic $\alpha$-effect
and quench the LSD such that the exponential growth of the large-scale
field transits to a resistively limited growth in the third stage.

\begin{figure}[h]
\includegraphics[width=0.48\textwidth]{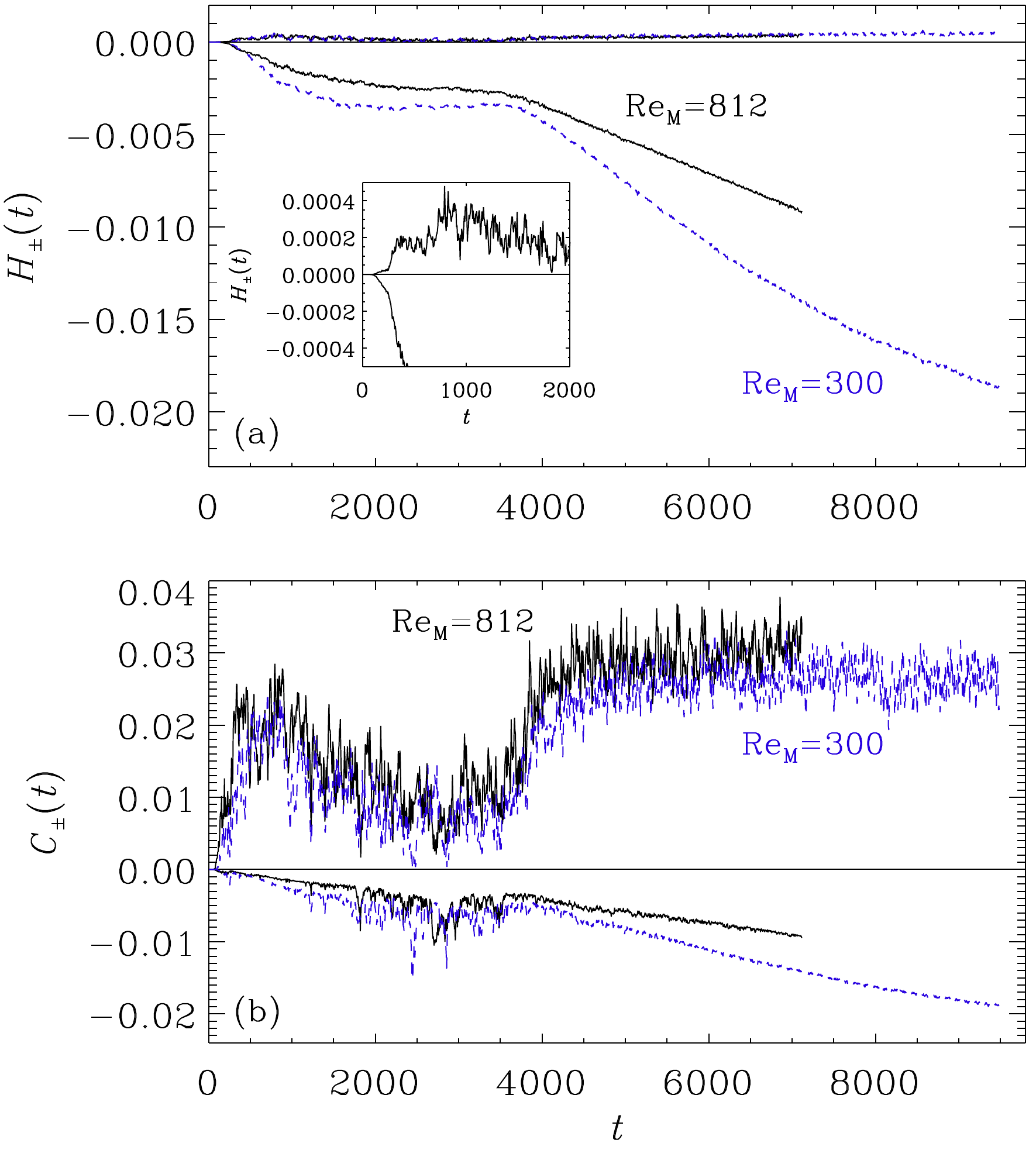}
\caption{Evolution of (a) $H_\pm(t)$ and (b) $C_\pm(t)$ for Run~A
and a similar run with $\Rm=300$.
The inset highlights the early growth of $H_+(t)$.
}\label{pchposneg}
\end{figure}

At late times, when the $\alpha^2$ dynamo is operating, magnetic helicity
continues to build up at larger scales; see \Fig{pchposneg}(a).
Here, $H_\pm(t)=\int_\pm H(k,t)\,dk$, where $\int_\pm$ denote the
integrals separately for positive and negative arguments, respectively.
The gradual build-up of $H_-(t)$ happens by dissipating magnetic helicity
of positive sign at small scales.
The dissipation of $H_+(t)$ is proportional to the corresponding
current helicity, $C_+(t)$, where $C_\pm(t)=\int_\pm k^2 H(k,t)\,dk$
are the contributions from positive and negative current helicity,
respectively.
Eventually, $C_+$ and $C_-$, begin to cancel each other;
see \Fig{pchposneg}(b).
This leads to the asymptotic steady state
where the total current helicity goes to zero and the
large-scale field goes to the box scale as in \cite{B01}.
Note this late time behaviour only occurs on the long 
resistive time scales.

\end{document}